\documentclass[journal]{IEEEtran}
\usepackage{cite}
\ifCLASSINFOpdf
  \usepackage[pdftex]{graphicx}
\else
  \usepackage[dvips]{graphicx}
  \DeclareGraphicsExtensions{.eps}
\fi
\usepackage{url}

\usepackage[cmex10]{amsmath}
\usepackage[tight,footnotesize]{subfigure}
\usepackage{fixltx2e}
\usepackage{stfloats}
\begin{document}
\title{A transposed frequency  technique for phase noise and frequency stability measurements }

\author{John~G.~Hartnett, Travis~Povey, Stephen~R.~Parker and Eugene~N.~Ivanov% <-this % stops a space
\thanks{John~G.~Hartnett, Travis~Povey, Stephen~R.~Parker and Eugene~N.~Ivanov are with the School of Physics, the University of Western Australia, Crawley, 6009, W.A., Australia. }% <-this % stops a space
\thanks{Manuscript received August 29, 2012; ....}}


\markboth{IEEE Trans. on Microwave Theory and Techniques,~Vol.~XX, No.~X, December~2012}%
{Shell \MakeLowercase{\textit{et al.}}: Bare Demo of IEEEtran.cls for Journals}



\maketitle

\begin{abstract}
The digital signal processing has greatly simplified the process of phase noise measurements, especially in oscillators, but its applications are largely confined to the frequencies below 400 MHz. We propose a novel transpose frequency technique, which extends the frequency range of digital noise measurements to the microwave domain and beyond. We discuss two basic variations of the proposed noise measurement technique, one of which enables characterization of phase fluctuations added to the passing signal by the particular microwave component, for example by an amplifier, while another one  is intended for measurements of phase fluctuations of the incoming signal itself.
\end{abstract}


% \fi
\IEEEpeerreviewmaketitle

\section{Introduction}
\IEEEPARstart{T}{he} Symmetricom test set has become a useful instrument for phase noise and frequency stability measurements that allows one to compare a signal from a device under test (DUT)  against a reference signal and as a result measure the relative phase noise and frequency stability between the two signals~\cite{Stein2010}. The manufacturer supplies a few models but currently the maximum input bandwidth  is 400 MHz (model 5125A) while the highest performance unit (model 5120-01), with the lowest phase noise measurement noise floor, only has a measurement bandwidth of 30 MHz. So how would one measure the additive or residual phase noise of a device under test which has an operating frequency much higher than 30 MHz or  even 400 MHz?

One method to achieve this is to take a beat note between two oscillators, and, by engineering a relatively low noise reference through frequency division of a signal derived from one of the oscillators, one can make a measurement on oscillators operating at X-band frequencies~\cite{Hartnett2012}. If the oscillators have similar phase noise 3 dB can be subtracted for their individual performance.  However this method is limited to the case where the beat note between the signals from two oscillators being compared falls within the measurement bandwidth of the test set being used.  

This paper outlines a method to use the Symmetricom test set outside its normal area of application (oscillator noise measurements). We show that it can be used for measuring phase fluctuations in passive and active microwave components (circulators, amplifiers, voltage controlled phase shifters etc) and at frequencies well outside its normal operational bandwidth. 

This method is quite different to that normally used for phase noise measurements where a homodyne technique uses baseband measurements and calibration of the mixer is required to calculate the phase noise from voltage noise measurements. It  allows one to simply measure the phase noise and/or frequency stability of practically any device at any frequency, provided the frequency is sufficiently stable. 

The DUT could be an active device where only phase noise is of interest or alternatively two oscillators where both phase noise and frequency stability are measured.

\section{Measurement technique}  
The essential elements of the measurement technique are shown in Fig. 1. Here the DUT is a microwave amplifier operating at frequency $f_0$. The microwave signal from a stable source is divided into two arms. One arm passes through the DUT and picks up the intrinsic phase fluctuations due to the amplifier. An attenuator (att1) is adjusted to give the desired input power and a second attenuator (att2) is used to provide the suitable LO drive level for the first mixing stage.  At this point an auxiliary source is used to provide a  signal at a frequency $f_a$ that is within the range of the test set being used. The important feature of this technique is to impose the DUT phase noise onto the low frequency $f_a$ auxiliary  signal, which is done via the use of two mixing stages.

%that after the microwave signal passes through the DUT it is mixed with the signal at frequency $f_a$ in the first mixing stage.  

\begin{figure}[!t]
\centering
\includegraphics[width=3.0in]{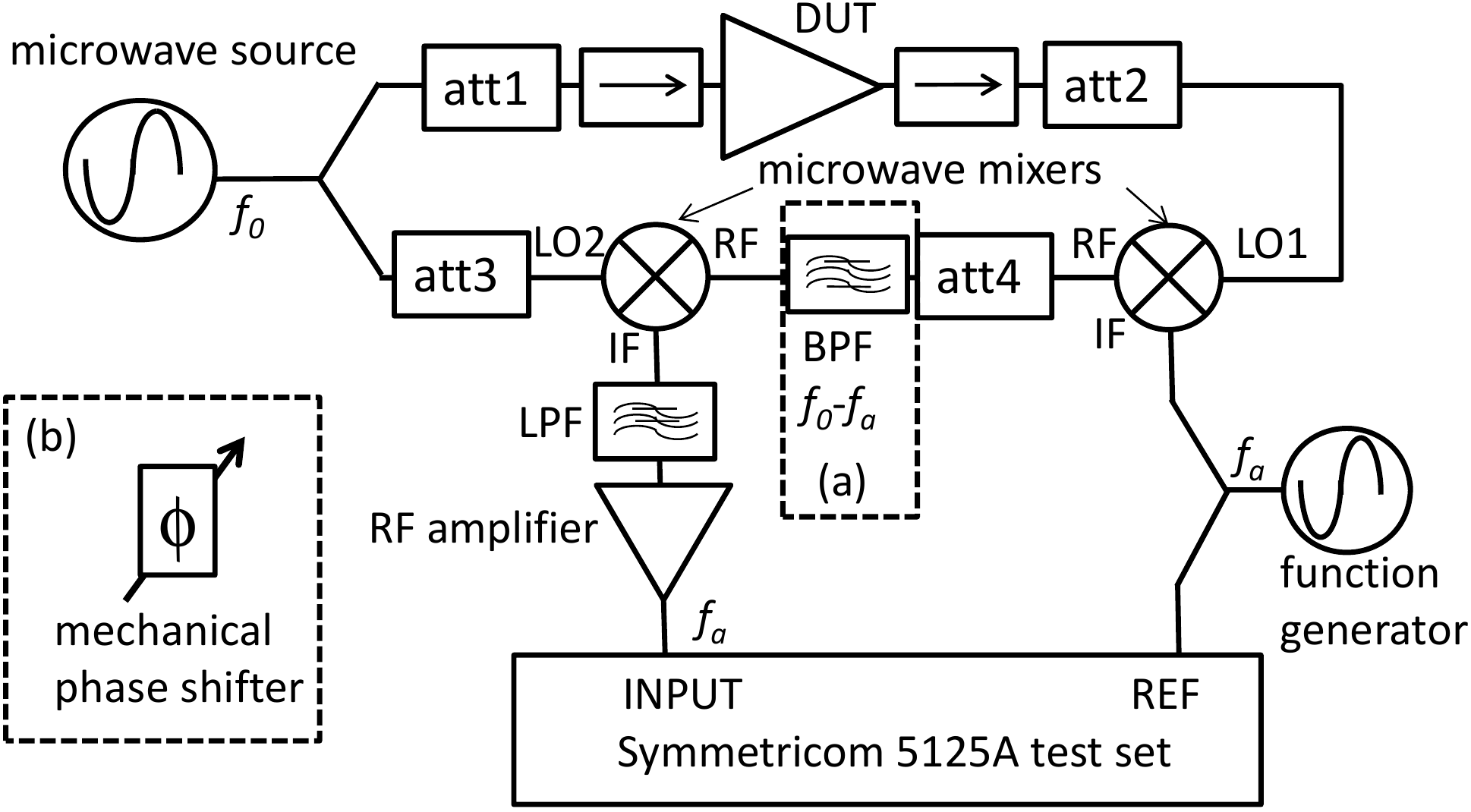}
\caption{A block diagram of the measurement system 1(a) and 1(b). DUT = Device Under Test. LPF = Low Pass Filter; BPF = Band Pass Filter;  attX = coaxial attenuator.  Both microwave synthesizer and the function generator were referenced by 10 MHz from a Kvarz CH1-75A hydrogen maser.  The dashed line rectangles indicate two implementations: 1(a) uses the BPF and 1(b) replaces the BPF with a (mechanical) phase shifter.}
\label{fig1}
\end{figure}

If necessary, attenuators (att3 and att4) are used to adjust the LO drive level and the RF power level to the second mixing stage. Alternatively an isolator could be used instead of attenuator att4. A band pass filter (BPF) is used to reject one of the sidebands, either $f_0 + f_a$ or $f_0- f_a$,  at the output of the first mixing stage. (An image rejection or single side band mixer could also be used.)  The second mixing stage recovers the  frequency ($f_a$) of the auxiliary signal, after low pass filtering (LPF) but it carries all the phase fluctuations of the device under test. The signal is then amplified and input to the Symmetricom test set. The reference to the test set is taken directly from the auxiliary oscillator.  This combination of two mixing stages with the test set is integral to this technique and allows one to essentially use the test set to measure the phase noise of any device operating at any frequency within the range of the components used.

\subsection{Measurement system with a band pass filter}

Considering the configuration of Fig. 1(a), we model the input signal to the second mixing stage, after filtering, as proportional to, $$cos[2 \pi (f_0 - f_a) t + \delta \varphi(t)+ \varphi_{LO1}],$$  where the upper sideband has been rejected,  and where $\delta \varphi(t)$ is the phase spectrum of the noise added by the DUT to the $f_0$ signal before being mixed with the $f_a$ auxiliary signal. The parameter  $\varphi_{LO1}$ represents the initial phase of the signal at LO1. After the second mixing stage  the resulting signal frequency dependence  can be modeled by the product of the  latter with the original $f_0$ microwave pump signal as, 
\begin{eqnarray}
&cos[2 \pi (f_0 - f_a) t + \delta \varphi(t)+ \varphi_{LO1}] cos[2 \pi f_0 t + \varphi_{LO2}] = \nonumber \\
&\frac{1}{2} cos[ 2 \pi f_a t + \delta \varphi(t)+ \Phi],
\label{eqn1}
\end{eqnarray}
where $\Phi = \varphi_{LO2}-\varphi_{LO1}$ is a phase constant and the high frequency $2 f_0$ mixing product has been rejected by the low pass filter. The $2 f_0$  signal is well suppressed by the mixer itself  if $f_0$ is in the microwave range. Here $\varphi_{LO2}$ represents the initial phase of the signal at LO2. This and the initial phase at LO1 are a fixed phase with respect to the initial phase of the auxiliary signal at the IF port of the first mixing stage. Therefore the last term in Eq. (\ref{eqn1}) is a constant phase factor and as a result  the $f_a$ signal input to the test set contains the phase noise of the DUT. 

The correlated phase fluctuations of the pump microwave signal ($f_0$) are rejected at the level of the second mixer. Phase fluctuations of the auxiliary oscillator are rejected by the test set because it measures fluctuations of the phase difference between two input signals (it is essentially a ``digital'' phase bridge).  The measurement system white noise floor is then determined primarily by the final low frequency RF amplifier used to provide sufficient signal power to the input of the test set.

The two mixers down converting the microwave signal, that passes through the DUT, to the frequency of the auxiliary signal, and the rejection of one of the sidebands, are the main elements in this technique. All variations on this design contain this important aspect. Ideas similar to this have been developed before under the name of ``transposed gain'' where a signal is down converted from a much higher microwave frequency to a low frequency where the amplifier, used as the important gain stage, has much lower intrinsic noise and after which the signal is up converted again to the higher microwave frequency~\cite{Driscoll1995, Everard1995a, Everard1995b, Everard1995c}. This technique is quite different as it does not actually use a transpose gain but transfers the noise of the DUT to a much lower frequency signal that is accessible to the phase noise test set used to measure the device's phase noise and/or frequency stability.

%The frequency of the auxiliary oscillator is chosen to suite the components used in the measurement system which are also within the bandwidth of the higher frequency pump oscillator driving the device under test. 

It is worth noting that the  DUT could have, alternately, been located on the other branch of the power splitter before the mixer input LO2 but we would get the same result as Eq. (\ref{eqn1}).

Caution needs to be taken to ensure that the LO power on the mixers is maintained at their optimum drive level and at the same level for all measurements. One might be measuring the DUT as a function of input power, as in Fig. 1, which will change the power at LO1. The mixer's contribution to the measurement system noise floor depends on its input power. It was observed that when the mixer power was allowed to drop well below optimum drive levels the measurement system noise floor was elevated and limited the measurement. 

\subsection{Measurement system without a band-pass filter}

However as indicated in Fig. 1(b) the band-pass filter could also be replaced with a (mechanical) phase shifter but it will be shown that this does not result in the correct measurement of the phase noise of the DUT. 

We model this configuration with a signal at the first mixing stage as proportional to, $$cos[2 \pi (f_0 \pm f_a)t + \delta \varphi(t)+ \varphi_{LO1} + \varphi_p],$$ where both sidebands are present, the term $\delta \varphi(t)$ represents the flicker phase fluctuations of the DUT. (The treatment of the white phase noise of the DUT is given below).  The term $\varphi_p$ represents the added phase from the variable (mechanical) phase shifter. When this is mixed at the second mixing stage the resulting signal frequency dependence  can be modeled by the product of the  latter with the original $f_0$ microwave pump signal as,
\begin{eqnarray}
&cos[2 \pi (f_0 \pm f_a)t + \delta \varphi(t) + \varphi_{LO1} + \varphi_p] \times \nonumber \\
&cos[2 \pi f_0 t + \varphi_{LO2} ] = \nonumber 
\\
%&\frac{1}{2}cos[2 \pi (-f_a) t + \delta \varphi(t)+\varphi_{LO1}+\varphi_p -\varphi_{LO2}] + \nonumber\\
%&\frac{1}{2}cos[2 \pi (f_a) t + \delta \varphi(t)+\varphi_{LO1}+\varphi_p  -\varphi_{LO2}]= \nonumber\\
%&\frac{1}{2}cos[2 \pi (f_a) t - \delta \varphi(t)-\varphi_{LO1}-\varphi_p +\varphi_{LO2}] + \nonumber\\
%&\frac{1}{2}cos[2 \pi (f_a) t + \delta \varphi(t)+\varphi_{LO1}+\varphi_p -\varphi_{LO2}]=\nonumber\\
&cos[2 \pi f_a t ]cos[\delta \varphi(t)- \Phi+\varphi_p],
\label{eqn2}
\end{eqnarray}
where the high frequency $2 f_0$ mixing product has been low pass filtered out and the phase noise sidebands are added to the phase constants. In contrast to the previous case, shown by Eq. (1), the phase noise of the DUT is not imposed on the output signal at frequency $f_a$. Therefore, the measurement system without the band-pass filter (Fig. 1(b)) is incapable of measuring \textit{flicker} phase fluctuations of the DUT. 

%By tuning the phase shifter $\varphi_p \rightarrow \varphi_{LO2}- \varphi_{LO1}$ the amplitude of the $f_a$ signal input to the test set is maximized as the right hand cosine term approaches unity. This leaves only the original auxiliary signal at frequency $f_a$. Hence one must  reject either of the two sidebands after the first mixing stage (using the band-pass filter or an image rejection mixer) in order to correctly measure the phase noise of the DUT.

%The attenuator (att4) is used to adjust for optimum input power to the RF amplifier. The measurement system noise floor is limited by the phase noise of the final RF amplifier. This technique produced a measurement system noise floor at the limit of the test set's noise floor itself.   The MiniCircuits ERA-5+ amplifier has sufficiently low enough intrinsic phase noise at high Fourier frequencies to approach the measurement system white noise floor of the Symmetricom 5125A test set at about -163 dBc/Hz (on the 20 MHz carrier). The MiniCircuits ZFL-500LN+ amplifier limited the measurement white noise floor to about -158 dBc/Hz.

\subsection{Broadband white noise}

In this section we show how the measurement system in Fig. 1(b) treats white phase noise of the DUT. In this case the output signal voltage from the DUT with effective temperature $T_{eff}$ as given by,
\begin{equation}
u_{in} = U_1 |T| cos[2 \pi f_0 t + \delta \varphi(t)+ \varphi_{LO1}] + \nu(t),
\end{equation}
where $U_1$ is the amplitude of the signal input to the DUT, $|T|$ is the DUT transmission coefficient and $\nu(t)$ is the white noise from the DUT with power spectral density,
\begin{equation}
S_{\nu} =k_B T_{eff} \, \,  \verb1[W/Hz], 1
\end{equation}
where $k_B$ is Boltzmann's constant. Therefore the power spectral density of phase fluctuations due to this thermal noise source  is,
\begin{equation}
S_{\phi} = \frac{k_B T_{eff}}{2 P |T|^2},
\end{equation}
where $P$ is the power of the DUT input signal.

After the reference phase shifter is adjusted to maximize the output signal from the second mixer, the input signal to the Symmetricom test set, can be modeled by,
\begin{equation}
u_{out} = \chi U_1 |T| cos[2 \pi f_a t ]+ \xi(t),
\label{eqn3}
\end{equation}
where $\chi$ is the mixer's conversion coefficient. The first term in Eq. (\ref{eqn3}) is the same as Eq. (\ref{eqn2}) but with the phase shifter tuned to maximize the $f_a$ signal, and the second term is the broadband white phase noise term, which unlike the first term is not dependent on the chosen reference phase shift. The power spectral density of $\xi(t)$ can be expressed as, 
\begin{equation}
S_{\xi} = \chi^2 k_B T_{eff} \,\, \verb1[W/Hz]. 1
\label{eqn3-2}
\end{equation}
The effective power spectral density of phase fluctuations of the output signal due to thermal noise is then, 
\begin{equation}
S_{\phi} = \frac{k_B T_{eff}}{2 P |T|^2}.
\end{equation}
This expression results from combining Eqs (\ref{eqn3}) and (\ref{eqn3-2}). It is power dependent and models what is observed. Equations (8) and (5) coincide. This proves that the measurement system in Fig. 1(b), while insensitive to flicker phase noise of the DUT, can correctly measure its white phase noise. Hence the technique using the phase shifter, though it \textit{does not} allow us to measure the DUT narrow band phase noise on the auxiliary signal, it \textit{does correctly} measure its broadband white noise.

\begin{figure}[!t]
\centering
\includegraphics[width=3.0in]{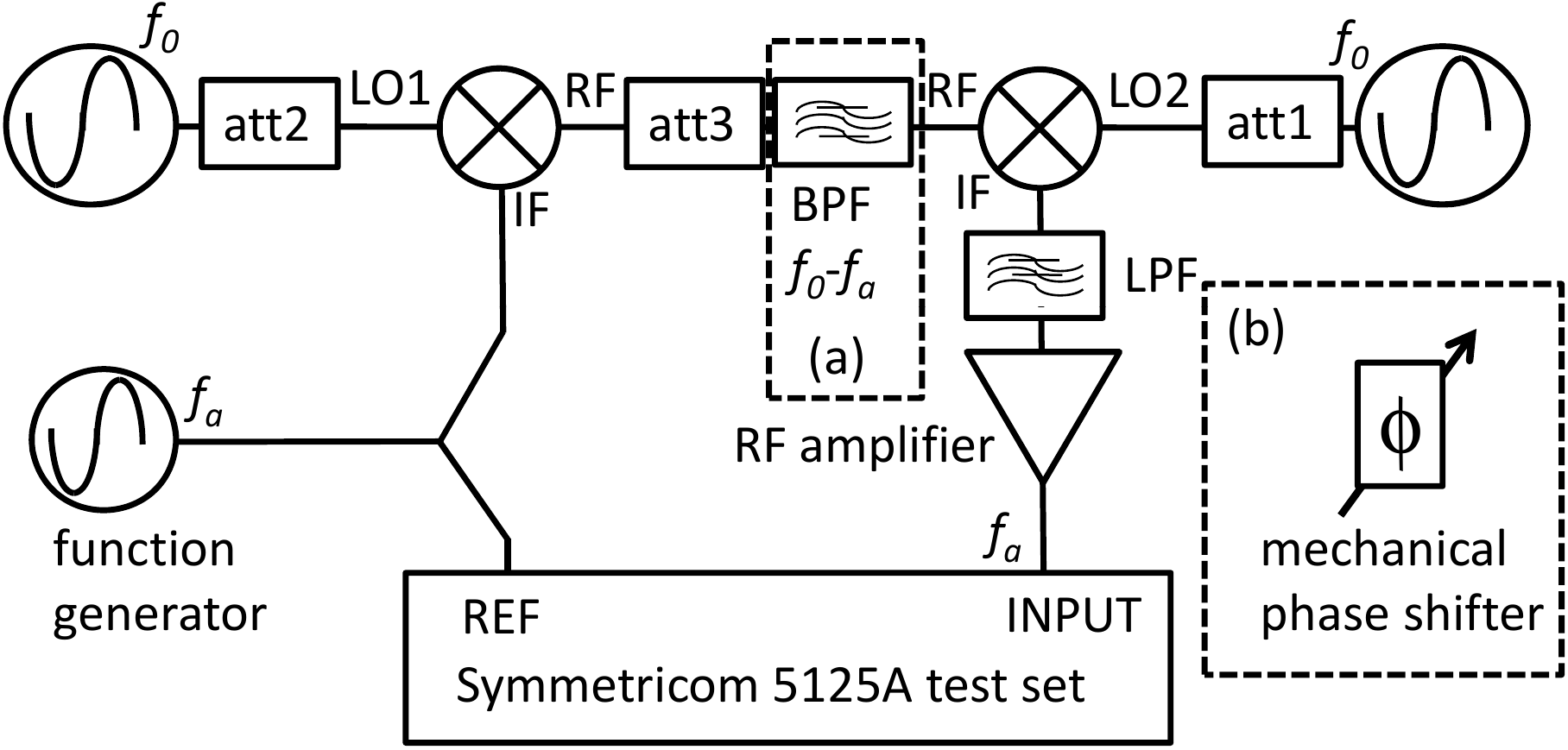}
\caption{A block diagram of the measurement system used to measure the intrinsic phase noise of oscillators. The same key elements are employed as in Fig. 1(a): the two mixing stages and the band-pass filter that down convert the frequency of the oscillator under test to that of the auxiliary oscillator. A band-pass filter is essential between the two mixers.  
}
\label{fig2}
\end{figure}

\subsection{Oscillator comparisons}

The same key elements of the configuration of Fig. 1(a), using the band-pass filter, can be used to measure the phase noise and frequency stability of a pair of  oscillators at different frequencies, where we assign $f_{01}$ and $f_{02}$ to the frequencies of the two oscillators. The same aspects apply to this configuration as discussed above, so we'll only consider the situation where a band-pass filter is used. If the phase shifter is used instead of the band-pass filter it can be shown, like in Eq. (2), that the relative phase fluctuations of the  oscillators are not recovered on the auxiliary signal.

Fig. 2 describes the basic features for the measurement method for two oscillators where a band-pass filter is used after the first mixing stage. The main difference with Fig. 1 is that two devices are needed. One may be used as a DUT and the other as a reference to obtain the measured signal frequency where, 
\begin{equation}
f_a \pm  |f_{01} - f_{02}| < 400 \,\, \verb1MHz, 1
\label{eqn400}
\end{equation}
with the $\pm$ depending on the choice of band-pass filter.  This ensures a signal at a frequency measureable by the test set containing the phase fluctuations of the two oscillators. Here we have assumed a 400 MHz bandwidth test set is being used. Therefore Eq. (\ref{eqn400}) means the frequency difference between the oscillators  must be less than twice the bandwidth of the test set otherwise additional mixing stages are needed.

In the case where a reference oscillator has much lower phase noise than the DUT oscillator the final result is that of the DUT but if they are nominally identical oscillators then we get the relative phase fluctuations of the two devices. This means 3 dB must be subtracted for the phase noise of a single oscillator.

From one of the oscillators an intermediate frequency is generated with an auxiliary oscillator in the first mixing stage, which is filtered to reject one of the sidebands with the band-pass filter, and then mixed in the second stage with the signal from the other  oscillator, low pass filtered, amplified and measured on the Symmetricom test set.   The phase noise is read directly from the test set, but to find the Allan deviation of the signal at frequency $f_{01}$ or $f_{02}$ one must scale by the ratio $(f_a \pm |f_{01}-f_{02}|)/f_{0}$ where $f_0$ is chosen as equal to either $f_{01}$ or $f_{02}$. 

In the case of Fig. 2 we can model the results of the first mixing stage as proportional to, $$cos[2 \pi (f_{01} - f_a)t + \delta \varphi_1 (t)+ \varphi_{LO1}],$$ where upper $f_a$ sideband has been filtered out by the band-pass filter. The phase noise of the oscillator is represented by $\delta \varphi_1 (t)$. When this is mixed at the second mixing stage with the output of the second oscillator with its own phase noise $\delta\varphi_2 (t)$ at the  frequency $f_{02}$, and modeled as proportional to, $$cos[2 \pi f_{02} t +\delta \varphi_2 (t) + \varphi_{LO2} ],$$ the resulting signal frequency dependence can be modeled by the product of the  latter with the former as,
\begin{eqnarray}
&cos[2 \pi (f_{01} - f_a)t +\delta \varphi_1(t) + \varphi_{LO1}] \times \nonumber \\
&cos[2 \pi f_{02} t +\delta \varphi_2 (t) + \varphi_{LO2} ] = \nonumber \\
&\frac{1}{2}cos[2 \pi (f_{01}-f_{02}+f_a) t +\delta \varphi_1(t)+\delta \varphi_2 (t) + \Phi],
\label{eqn4}
\end{eqnarray}
where $\Phi = \varphi_{LO2}-\varphi_{LO1}$ is a phase constant and the high frequency  mixing product of order $f_{01}+f_{02}$ has been filtered out. The power of phase fluctuations of the resulting signal at frequency $f_{01}-f_{02}+f_a$ is equal to the  combined power of phase fluctuations of the individual microwave oscillators.

Of course the low pass filter must now pass the signal with frequency $f_{01}-f_{02}+f_a$, and this frequency must be within the bandwidth of the measurement test set. This then necessarily affects the frequency at which the phase noise is measured and the scaling ratio to calculate the Allan deviation becomes $(f_{01}-f_{02}+f_a)/f_{0}$. If the oscillators are nominally identical an additional $1/\sqrt2$ factor must be applied to get the stability of a single oscillator.

\begin{figure}[!t]
\centering
\includegraphics[width=3.0in]{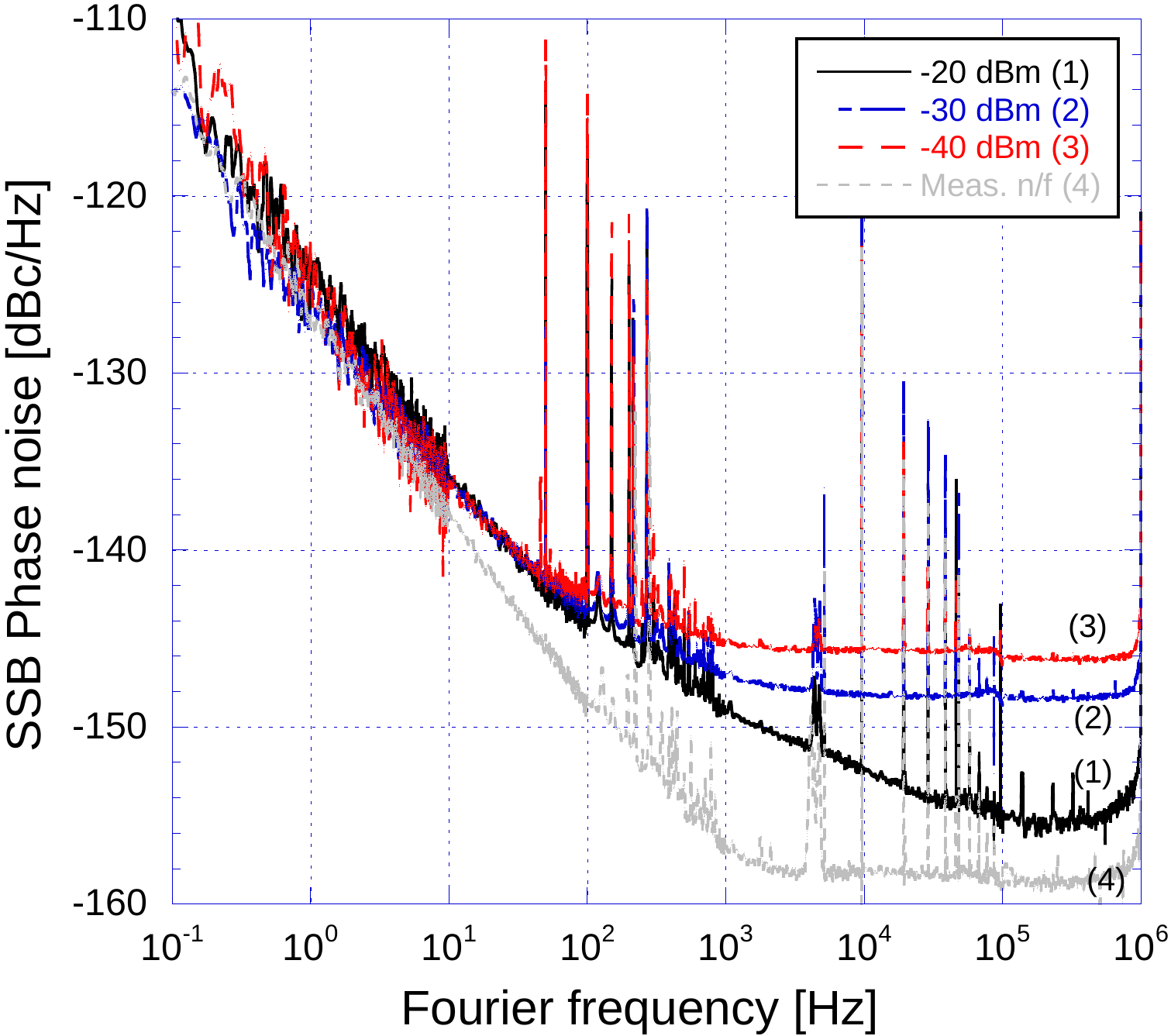}
\caption{The SSB phase noise, as a function of input power of the Endwave amplifier model no. JCA812-5001, operating at 11.2 GHz, measured  using the measurement method  of Fig. 1(a) with $f_a = 20$ MHz. The legend indicates the input power to the amplifier and the measurement noise floor (curve 4) which was measured by removing the DUT from the circuit. RF amplifier = MiniCircuits ZFL-500LN+.}
\label{fig3}
\end{figure}

\begin{figure}[!t]
\centering
\includegraphics[width=3.0in]{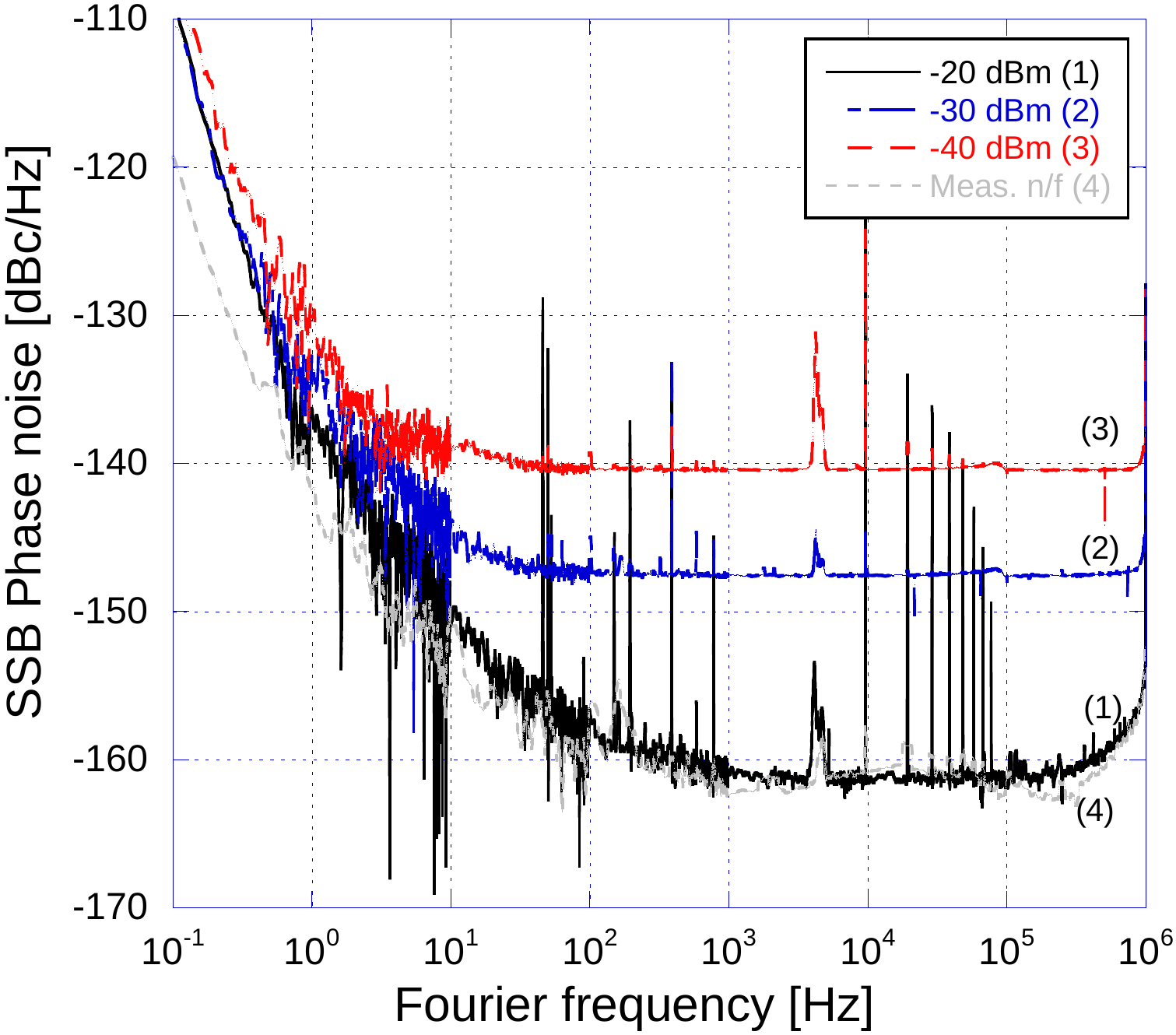}
\caption{After replacing the BPF with a mechanical phase shifter (Fig. 1(b)) the SSB phase noise of the output signal at $f_a$ = 20 MHz was measured, where a AML amplifier model no. AML812PNA5402, operating at 11.2 GHz, was used with different input power levels. The legend indicates the input power to the amplifier and the measurement noise floor (curve 4) which was measured by removing the DUT from the circuit. A smoothing by taking a 50 point average was applied to data of curve 4. RF amplifier = MiniCircuits ERA-5+. See discussion in text.}
\label{fig4}
\end{figure}

\begin{figure}[!t]
\centering
\includegraphics[width=3.0in]{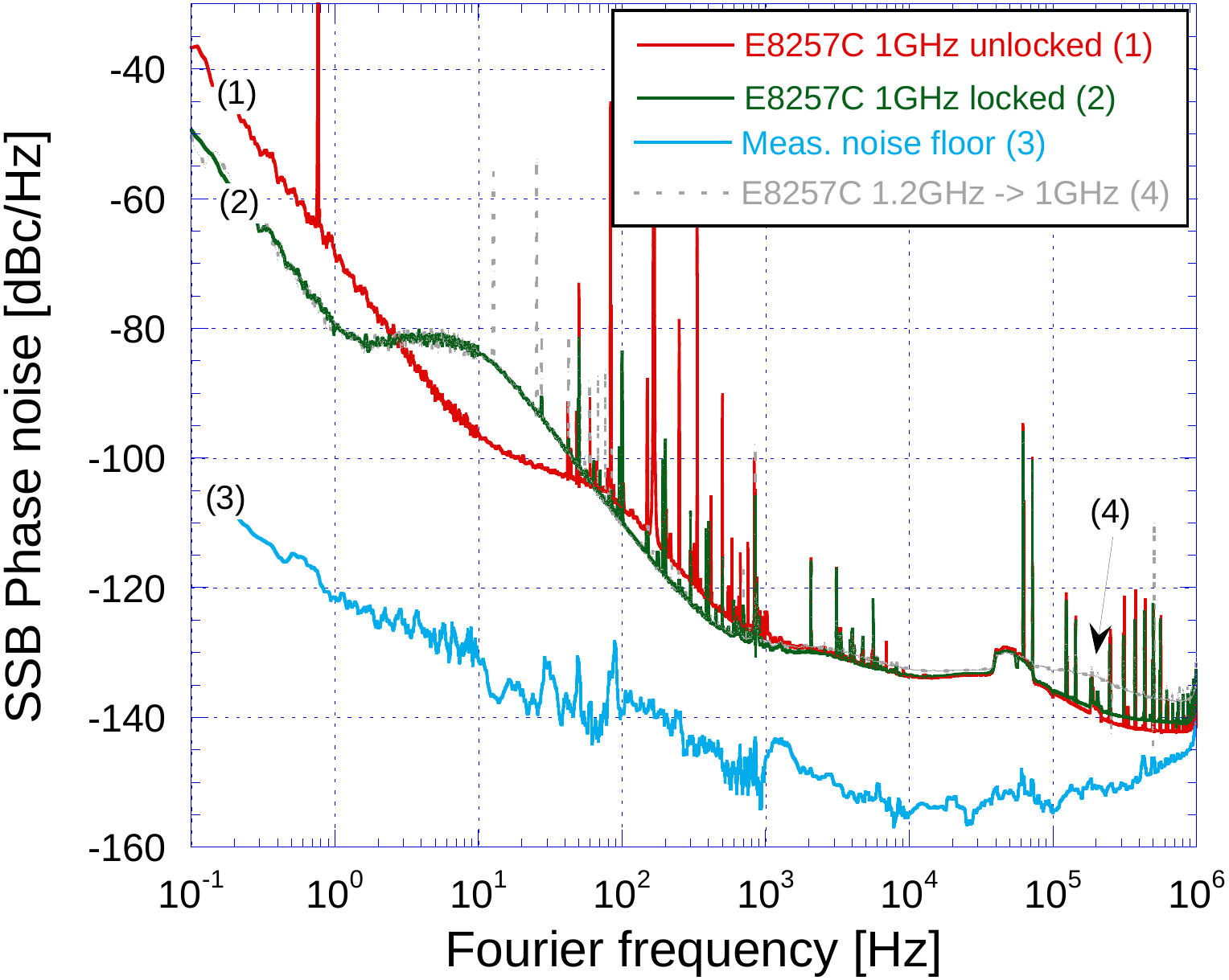}
\caption{The SSB phase noise of an Agilent E8257C synthesizer at 1 GHz without external reference (curve 1), and phase locked to a 10 Mz signal from the frequency doubled output of an Oscilloquartz 8607 quartz oscillator (curve 2). The latter was compared with a 1 GHz signal derived from an ultra-low phase noise cryogenic sapphire oscillator and measured using the measurement method of Fig. 2(a).  Curve 3 is the measurement system noise floor. Curve 4 is the phase noise of the  locked synthesizer at 1.2 GHz but referred to 1 GHz. See text for details.}
\label{fig5}
\end{figure}

\section{Results}

Using the measurement method of Fig. 1(a) we measured the phase noise of an Endwave amplifier (model number JCA812-5001) as a function of input power. The band-pass filter is tunable home made cavity filter and the low-pass filter a  MiniCircuits SLP-21.4+.  The microwave mixers are Watkins Johnson model number M14A. The function generator is a 20 MHz Agilent model 33220A. The final RF amplifier used was either a MiniCircuits ZFL-500LN+ or an ERA-5+ amplifier.

The results are shown in Fig. 3. Where these are above the measurement noise floor, they are consistent with those expected for the amplifier and indicate a flicker law for Fourier frequencies $10 \leq f \leq 100$ Hz. The thermal noise floor is dependent on input power to the amplifier as expected. A carrier frequency of $f_0 = 11.2$ GHz and an auxiliary  frequency  $f_a = 20$ MHz were used in these measurements. 

The measurement system noise floor was measured by removing the DUT and it was found that it limits the measurements for Fourier frequencies $f < 10$ Hz. The phase noise of the final RF amplifier was determined to be the principle limitation of the measurement system especially at Fourier frequencies $f > 100$ Hz. By using either a MiniCircuits ERA-5+ or a ZFL-500LN+ amplifier the measurement system white noise floor was limited to about -163 dBc/Hz and about -158 dBc/Hz  (on the 20 MHz carrier) respectively. The former is very close to white noise floor of the Symmetricom 5125A test set itself. 

The additive phase noise of the RF amplifiers used here were separately measured by splitting the 20 MHz signal of the auxiliary oscillator and passing one signal through the amplifier to the input channel of the test set and the other directly to the reference channel. In both cases, the same input power to the amplifier was used as in the measurement system, and the results matched the thermal noise floor of the measurement system when using these RF amplifiers. 

Finally, when using the ERA-5+ RF amplifier,  a pair of MiniCircuits ZX05-153+ mixers were substituted for the Watkins Johnson M14A mixers and this raised the measurement system white noise floor 3 dB to about -160 dBc/Hz, indicating that a small contribution also may come from the choice of mixers used.

Using the method of Fig. 1(b) and  a very low phase noise AML X-band microwave amplifier we measured essentially the phase noise floor of the measurements system as a function of input power to the X-band amplifier. In this configuration where a mechanical phase shifter (an Arra 9428B in this case) is used instead of the band-pass filter one does not get the phase noise of the DUT (nor that for the oscillator(s) if a phase shifter is used in the configuration of Fig. 2). The  results are shown in Fig. 4. However the measurements of Fig. 4 indicate that as one decreases the input power to the X-band amplifier (the DUT) one observes rising thermal noise. At input power levels $\geq -20$ dBm (curve 1) the measured noise is at the measurement system noise floor. At input power levels less than this we observe an increase in white noise approximately proportional to the change in input power. 

The measurement system noise floors for both configurations shown in Fig. 1 were determined by removing the DUT, and ensuring the power to the mixers and the RF amplifier were the same as used when making the DUT phase noise measurements. Looking at Figs 3 and 4 there is a clear difference in the frequency dependence and level of the measurement system phase noise floor. The difference is that in the method of Fig 1.(a), with the band-pass filter, the mixer noise is present for $f < 10^3$ Hz (curve 4 in Fig. 3) but in the method of  Fig 1.(b), without the band-pass filter, the mixer noise is not present (curve 4 in Fig. 4) and we only measure the test set noise floor for $f < 10^2$ Hz.  At higher offset frequencies it is the RF amplifier that sets the thermal noise floor.

As shown in Eq. (\ref{eqn2}), the DUT phase noise is not added to the auxiliary signal, when the phase shifter was substituted for the band-pass filter, so also the mixers in the circuit are devices under test and their phase noise is not added but shows up as amplitude noise on the signal input to the test set. By contrast, from Eq. (\ref{eqn1}), when the band-pass filter is used, the  phase noise of the mixers is, in addition to the phase noise of the DUT, added to the signal input to the test set. Therefore with the DUT removed we measure a system noise floor that includes the phase noise of the two mixers and the RF amplifier (at the higher offset frequencies).

\begin{figure}[!t]
\centering
\includegraphics[width=3.0in]{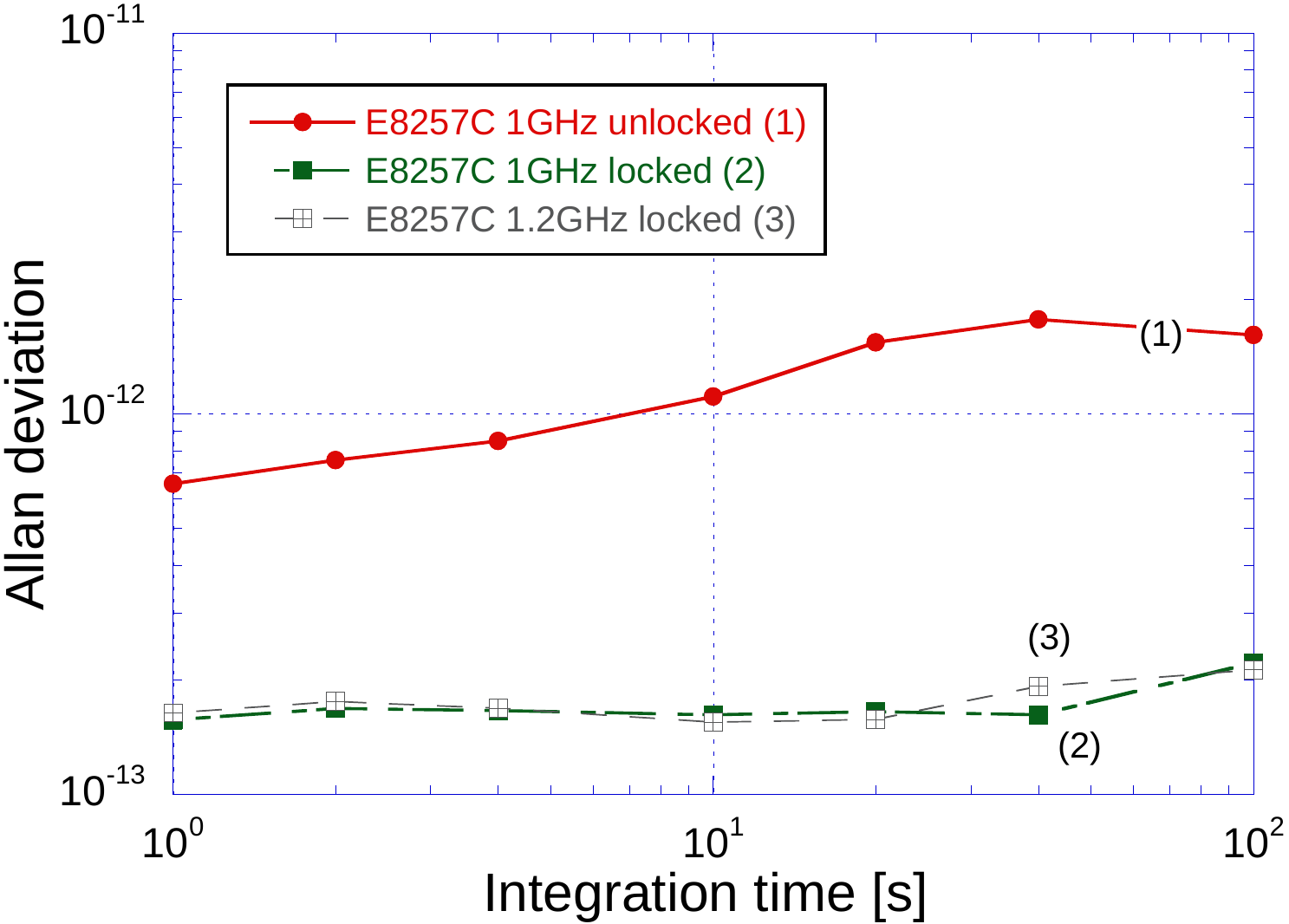}
\caption{Allan deviation of an Agilent E8257C synthesizer at 1 GHz without external reference (curve 1), and phase locked to a 10 Mz signal from the frequency doubled output of an Oscilloquartz 8607 quartz oscillator (curve 2). The latter was compared with a 1 GHz signal derived from an ultra-low phase noise cryogenic sapphire oscillator and measured using the measurement method of Fig. 2(a). Curve 3 is the Allan deviation of the signal frequency of the locked synthesizer at 1.2 GHz after the appropriate scaling was applied. See text for details.}
\label{fig6}
\end{figure}

Using the measurement method of Fig. 2 we measured the phase noise of an Agilent E8257C synthesizer generating a 1 GHz signal compared to a 1 GHz signal synthesized from an ultra-low phase noise cryogenic sapphire oscillator. The phase noise spectra are shown in Fig. 5, and the frequency stability in Fig. 6. Curve 1 represents the phase noise of the synthesizer without external reference, and curve 2 when it was referenced by a 10 Mz signal from the frequency doubled output of an Oscilloquartz 8607 quartz oscillator. 

The measurement system used two MiniCircuits ZX05-10L-S+ mixers, a ZVBP-909-S+ band-pass filter, a SLP-100+ low pass filter and a ZFL-500LN+ amplifier. Its measurement system phase noise floor is shown in curve 3 in Fig. 5. This was determined by using the same low phase noise oscillator on both input ports (for Oscillators 1 and 2 in Fig. 2). The auxiliary oscillator frequency used in these measurements was $f_a = 100$ MHz, derived from the  ultra-low phase noise cryogenic sapphire oscillator, with a phase noise of -130 dBc/Hz at 1 Hz offset  \cite{Nand2010, Hartnett2012}. In this case it was necessary to scale resulting stability generated by the test set by $f_a/f_{0} = 100$ MHz$/1.0$ GHz $= 1/10$ where $f_{0} = f_{01}= f_{02} = 1$ GHz.

We also compared the same oscillators where we raised the output signal frequency of the  Agilent  E8257C synthesizer to $f_{02}=1.2$ GHz.  The auxiliary oscillator was still at $f_a = 100$ MHz, which was mixed with the signal at $f_{01} =1$ GHz from the ultra-low phase noise cryogenic sapphire oscillator, then we filtered out the upper sideband with the band-pass filter  leaving the lower sideband at 900 MHz to be mixed with the 1.2 GHz signal. After low-pass filtering (using a MiniCircuits SLP-400 filter) and amplification this resulted in a 300 MHz signal  that was measured by the test set. Curve 4 in Fig. 5 is the result where a factor of $20 log(1.2)$ has been subtracted to compare the results all at 1 GHz. It is essentially identical with curve 2 as expected.  And in Fig 6 we show the Allan deviation of the 1.2 GHz synthesizer signal where the correct scaling has been applied to the output of the test set.  In this case $(f_{01}-f_{02}+f_a)/f_{02} = 300$ MHz$/1.2$ GHz $= 1/4$. Only Allan deviation data with a noise equivalent bandwidth (NEQ) of 0.5 Hz is shown.

The cryogenic sapphire oscillator's phase noise and frequency stability \cite{Hartnett2012} are orders of magnitude lower than that of the Agilent E8257C synthesizer. At 1 GHz its phase noise is approximately equal to the measurement system noise floor (curve 3 or Fig. 5), hence it does not contribute to these results. Figs 5 and 6 therefore show only the performance of the  E8257C synthesizer.

It should be noted that when the frequency difference of the two oscillators falls within the bandwidth of the test set the additive phase noise can be measured by taking the beat note of the two signals and comparing it to a previously characterized low noise reference. To confirm the validity of our transposed gain technique we used this direct comparison method to repeat the measurements in figures 5 and 6 and found no discrepancy in the results.

%\section{Discussion}

%In order to measure an extremely low phase noise amplifier, for example, with a thermal noise floor of, say, -175 dBc/Hz, one would have to use the Symmetricom 5120A-01 test set with ultra-low-noise option. Also one would need to make sure that the RF amplifier, required after the second mixing stage,  has phase noise lower than this level. The phase noise of the RF amplifier depends on its input power which is set by the conversion efficiency of the second mixer.

%In the case of the oscillator measurements at 1 GHz (using Fig. 2) the phase noise floor seen in Fig. 5 is determined largely by the RF amplifier. The ZX05-10L-S+ mixers, the ZVBP-909-S+ band-pass filter and a ZFL-500LN+ amplifier comprised the measurement system here.  Initially we used the 100 MHz from our hydrogen maser as the auxiliary signal, but found it limited the noise floor of the measurement system to -140 dBc/Hz due to only 20 dB rejection by the mixers of the common mode phase noise and a strong peak near 5 kHz which could be seen in the results. This is apparent in the phase noise comparison of the synthesizers at 11.2 GHz in Fig. 7. By using the 100 MHz synthesized from the ultra-low noise sapphire oscillator  instead this effect was eliminated.   

\section{Conclusion}
A technique has been developed that allows one to measure the phase noise of active components operating at frequencies well outside the bandwidth of the phase noise measurement test set being used. 

Digital measurements of phase fluctuations offer several advantages over the usual homodyne technique. Here is no need for the calibration of the mixer nor is there any requirement to  phase lock two oscillators to maintain zero voltage at the output of the mixer during the measurement process. The latter feature  permitted, for the first time, the phase noise measurement of ultra-stable cryogenic sapphire oscillators. 

This method is only limited by the operational bandwidth of the components used, and the noise floor imposed on the measurements by the two mixing stages and the final RF amplifier used to provide sufficient signal power to the test set. Using the band-pass filter and X-band mixers a measurement system noise floor at low offset frequencies of $10^{-12.7}/f$ was obtained, while with 1 GHz mixers $10^{-12.0}/f$ was achieved. Using a MiniCircuits ERA-5+ RF amplifier at the input to the test set a measurement system  white noise floor of -163 dBc/Hz was obtained.

Hence one may achieve a measurement white noise floor at the limits of the test set itself, near -165 dBc/Hz for the 5125A model and -175 dBc/Hz for 5120A-01 model with ultra-low-noise option. Provided the final RF amplifier is sufficiently low noise one can reach this white noise floor and potentially measure the phase noise of extremely low noise components.

\section{Acknowledgments}
This work was supported by ARC grant LP110200142.

\end{document}